\pdfoutput=1
\documentclass[10pt,aps,prl,showpacs,amsmath,amssymb,
							 twocolumn,raggedbottom,superscriptaddress]{revtex4-1}
\usepackage[bookmarks=false]{hyperref}
\usepackage{amsmath}
\usepackage{amssymb}
\usepackage{bm}
\usepackage{amsthm}
\usepackage{amsfonts}
\usepackage{latexsym}
\usepackage{graphicx}
\usepackage{microtype}

\begin{document}

\title{Asymmetric orbital-lattice interactions in ultra-thin correlated oxide films}

\author{J.\! Chakhalian}
\email{jchakhal@uark.edu}
\affiliation{%
Department of Physics, University of Arkansas, Fayetteville, Arkansas 70701, USA}
\author{J.M.\! Rondinelli}
  \affiliation{Department\! of\! Materials\! Science\! \&\! Engineering,\! 
	Drexel\! University,\! Philadelphia,\! Pennsylvania\! 19104,\! USA}
\affiliation{%
Advanced Photon Source, Argonne National Laboratory, Argonne, Illinois 60439, USA}
\author{Jian\! Liu}
\affiliation{%
Department of Physics, University of Arkansas, Fayetteville, Arkansas 70701, USA}
\affiliation{%
Advanced Light Source, Lawrence Berkeley National Laboratory, Berkeley, California 94720, USA}
\author{B.A.\! Gray}
\affiliation{%
Department of Physics, University of Arkansas, Fayetteville, Arkansas 70701, USA}
\author{M.\! Kareev}
\affiliation{%
Department of Physics, University of Arkansas, Fayetteville, Arkansas 70701, USA}
\author{E.J.\! Moon}
\affiliation{%
Department of Physics, University of Arkansas, Fayetteville, Arkansas 70701, USA}
\author{N.\! Prasai}
\affiliation{Department of Physics, University of Miami,  Coral Gables, Florida, 33124, USA}
\author{J.L.\! Cohn}
\affiliation{%
Department of Physics, University of Miami,  Coral Gables, Florida, 33124, USA}
\author{M.\! Varela}
\affiliation{%
Oak Ridge National Laboratory, Oak Ridge, Tennessee 37831, USA}
\author{I.C.\! Tung}
\affiliation{Materials Science and Engineering, Northwestern University, Evanston, Illinois, 60208, USA}
\author{M.J.\! Bedzyk}
\affiliation{Materials Science and Engineering, Northwestern University, Evanston, Illinois, 60208, USA}
\author{S.G.\! Altendorf}
   \affiliation{II.\ Physikalisches Institut, Universit{\"a}t zu K{\"o}ln,
    Z{\"u}lpicher Str.\ 77, 50937 K{\"o}ln, Germany}
\author{F.\! Strigari}
   \affiliation{II.\ Physikalisches Institut, Universit{\"a}t zu K{\"o}ln,
    Z{\"u}lpicher Str.\ 77, 50937 K{\"o}ln, Germany}
\author{B.\! Dabrowski}
\affiliation{%
Department of Physics, Northern Illinois University, Dekalb, Illinois 60115, USA}
\author{L.H.\! Tjeng}
    \affiliation{Max Planck Institute for Chemical Physics of Solids,
    N\"othnitzerstr.\ 40, 01187 Dresden, Germany}
\author{P.J.\ Ryan}
\affiliation{%
Advanced Photon Source, Argonne National Laboratory, Argonne, Illinois 60439, USA}
\author{J.W.\ Freeland}
\affiliation{%
Advanced Photon Source, Argonne National Laboratory, Argonne, Illinois 60439, USA}

\begin{abstract}\sloppy
Using resonant X-ray spectroscopies combined with density functional calculations, we find an asymmetric bi-axial strain-induced  $d$-orbital response in ultra-thin films of the correlated metal LaNiO$_3$ which are not accessible in the bulk. The sign of the misfit strain governs the stability of an octahedral ``breathing'' distortion, which, in turn, produces an emergent charge-ordered ground state with an altered ligand-hole density and bond covalency.  Control of this new mechanism opens a pathway to rational orbital engineering, providing a platform for artificially designed Mott materials.
\end{abstract}
\date{\today}
\pacs{73.20.-r, 73.50.-h, 68.55.-a, 71.20.Be}
\maketitle

Heteroepitaxial synthesis is a powerful avenue to modify  
orbital--lattice interactions in correlated  materials 
with strong electron--electron interactions derived from 
transition metals with open $d$-shell configurations \cite{Takagi/Hwang:2010,Mannhart/Schlom:2010,Hormoz/Ramanathan:2010}.
Epitaxial strain allows access to latent
electronic functionalities and  phases
that do not exist in bulk equilibrium phase diagrams 
\cite{%
Schlom/Fennie_etal:2010,%
Cao/Wu_et_al:2009,%
Hatt/Ramesh:2009,%
Wakabayashi/Sawa_et_al:2006,%
Okamoto/Millis:2004a}.
However, efforts to rationally control properties that 
are exceedingly sensitive to small perturbations \cite{Dagotto:2005}  
through the orbital--lattice interaction 
are impeded by the poor understanding of how
heteroepitaxy imposes 
constraints on the orbital response \cite{kiemer_et_al:2011,kiemer_et_al:2011b,Tokura/Nagaosa:2000}.
Despite the recent progress in strain-induced orbital engineering, 
a crucial fundamental question remains:
when a  single electron occupies a doubly degenerate $d$-orbital
in a cubic crystal field as in perovskites with Cu$^{2+}$, Mn$^{3+}$ or
low spin Ni$^{3+}$ cations, how does the substrate imposed epitaxial 
constraints dictate the correlated orbital responses of ultrathin films?
The exceptional strain control of the frontier atomic orbitals
 relies on the susceptibility of the orbital occupations
and their energy level splittings
to biaxial strain-induced lattice deformations, i.e.\ orbital--lattice 
interactions.
The conventional orbital engineering approach
in perovskite-structured oxides is often rationalized as follows:
coherent heteroepitaxy imposes a tetragonal distortion
on the film's primitive unit cell, which then 
modifies the chemical bond lengths of the 
functional octahedral building blocks.
The bond distortions in turn alter the crystal field symmetry 
and remove the cubic twofold $e_g$ electron degeneracy.
Since each $e_g$ orbital state is of the same symmetry, it is generally
anticipated that both tensile and compressive strains should
symmetrically alter the $d_{x^2-y^2}$ and $d_{3z^2-r^2}$ orbital states.
They are either lowered or raised relative to the strain-free band center of 
mass, which for finite filling, leads to an orbital polarization.
This symmetric strain-induced orbital polarization (SIOP) concept is 
routinely used to rationalize the orbital responses of many complex oxide systems \cite{Tokura/Nagaosa:2000,Ahh/Lookman/Bishop:2004}
and even  theoretically suggested to be
a possible route to replicate high-$T_c$ cuprate physics in nickelates  \cite{Chaloupka/Khaliullin:2008,*Hansmann/Held_et_al:2009,*Hansmann_et_al:2010}.
However, the strategies for tuning orbital ground states
in {\it ultrathin} films devised from the supposedly symmetric
strain response of the orbital occupations alone are
violated more often than they apply: Doped manganite thin
films exhibit $d_{3z^2-r^2}$ orbital polarization regardless of the bi-axial 
strain sign \cite{Khomskii/vandenBrink:2000,Tebano/Brookes_et_al:2008}. 
Ultrathin cuprate bilayers also show variable critical temperatures
 \cite{Gozer/Bozovic_etal:2008} correlated with the sign of the
interface lattice misfit \cite{Hua/Bozovic_et_al:2010},
and cobaltite films are either ferro- or diamagnetic depending on the
strain state \cite{Fuchs_et_al:2008}.
In this Letter, we  provide insight into why many {\it ultrathin} correlated oxides  violate the SIOP
model by investigating the orbital--lattice interactions in
10 unit cell thick films of the correlated and orbitally degenerate ($t_{2g}^6e_g^1$)
metal LaNiO$_3$ (LNO).
This representative spin-$\frac{1}{2}$ system has no magnetic or structural transitions below 500~K,  which  otherwise might obscure the investigation \cite{Lacorre/Torrance:1992,Torrance/Niedermayer_etal:1992,Zhou/Bukowski:2000}.
We identify the microscopic mechanism responsible
for the asymmetric strain-induced orbital response and 
show that it originates from latent instabilities of the bulk material, which are strongly enhanced due to the epitaxial constraints imposed 
by the heterointerface.
We further demonstrate deterministic control of the ligand-hole density 
and covalency through the asymmetric orbital--lattice response, enabling 
access to an emergent charge-ordered phase not attainable in the bulk.
For this purpose, we synthesized high quality epitaxial LaNiO$_3$ films under 
different strain states by pulsed laser deposition on (001)-oriented single crystal
LaAlO$_3$ (LAO, lattice mismatch of $-$1.08\%),
SrLaGaO$_4$ (SLGO, +0.31\%), SrTiO$_3$ (STO, +1.83\%),
and GdScO$_3$ (GSO, +3.6\%) substrates.
Synchrotron-based X-ray diffraction confirms the
LNO films are fully and homogeneously strained \cite{Note1}. 
Consistent with previous studies on rhombohedral perovskites~\cite{May/Rondinelli:2010,stanford:2011}, we find our films have lower (monoclinic) symmetry than that of 
the bulk.

\begin{figure}[b]
\centering
\includegraphics[width=0.98\columnwidth]{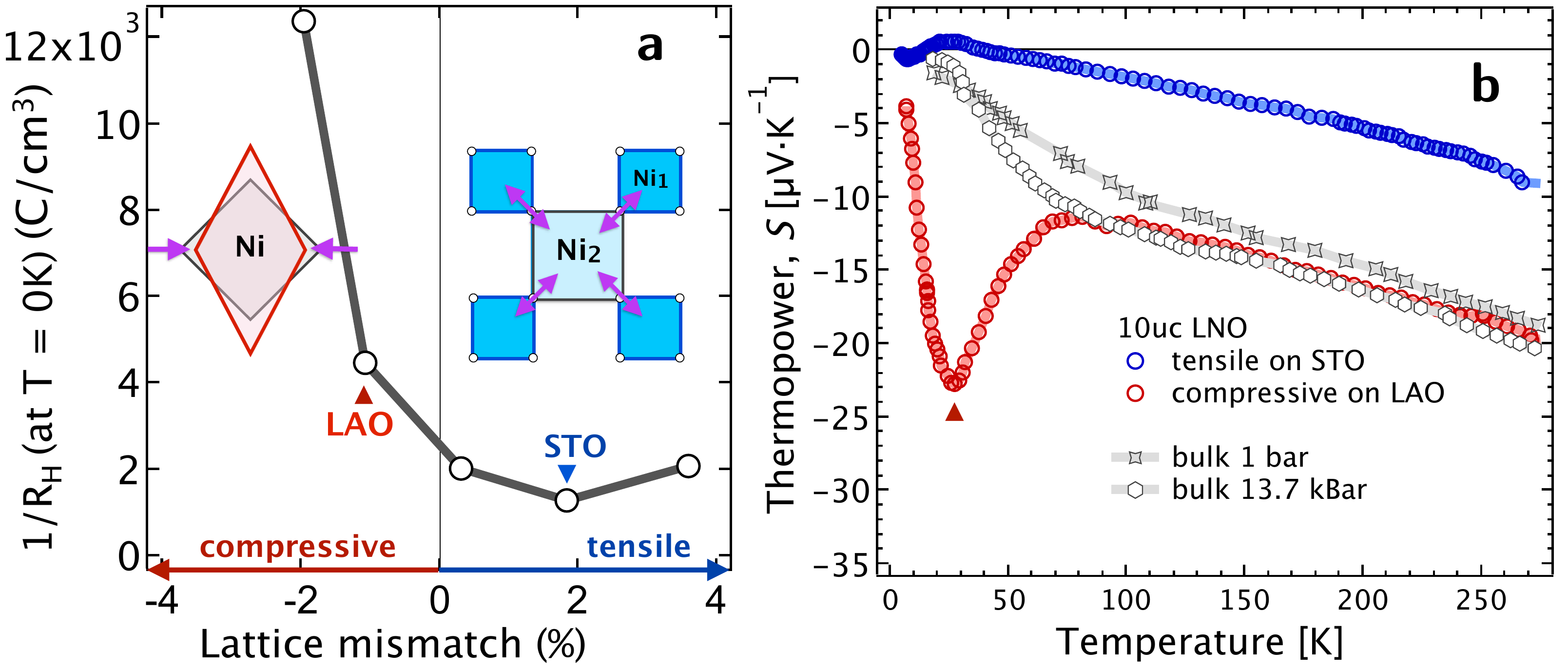}\vspace{-4pt}
\caption{\label{fig:RH_TEP}(Color online)
   Asymmetric strain-induced reponses: 
   (a) inverse Hall coefficient, $R_H^{-1}$ and 
   (b) temperature-dependent TEP.
   The carrier-density modulation in perovskite oxides is intimately
   linked with the the local structural of the NiO$_6$ octahedra (inset).
   The TEP in (b) reveals that epitaxial strain and hydrostatic  
   (bulk LNO under 1 bar and 13) kbar pressure are inequivalent. 
   Bulk data from Ref.~\cite{Zhou/Goodenough/Bukowski:2000}.}
\end{figure}
Figure~\ref{fig:RH_TEP}(a) reveals a  large anisotropy in the galvanoelectric 
response with the sign of the strain.
On the tensile side, the inverse Hall coefficient $R_H^{-1}$ is small
and independent of lattice mismatch indicating nearly zero strain-modulation in the carrier
density, while  for compressive strain an order of
magnitude increase in $R_H^{-1}$ occurs.
In addition, Fig.~\ref{fig:RH_TEP}(b) shows the thermoelectric power (TEP) over a broad temperature range
from 300~K down to 1.5~K for LNO films under different strain states.
The thermopower in films on the tensile side lack any noticeable anomalies and follow
the Mott relation \cite{Mott/Jones:1936}, though the magnitude is shifted relative
to the bulk toward positive values.  For compressive strain, the magnitude is
comparable to the bulk at T$>$80~K, but a different temperature
dependence emerges at low T: namely, a large negative peak,
reminiscent of phonon drag, occurs at 25~K. The asymmetric
strain-induced TEP response indicates distinct differences
in the electronic structure under tensile and compressive strain.
In contrast, hydrostatic pressure produces only modest enhancements in the TEP magnitude 
[Fig.~\ref{fig:RH_TEP}(b)], indicating compressive bi-axial strain is inequivalent to hydrostatic pressure.

\begin{figure}[b]
\centering
   \includegraphics[width=0.96\columnwidth]{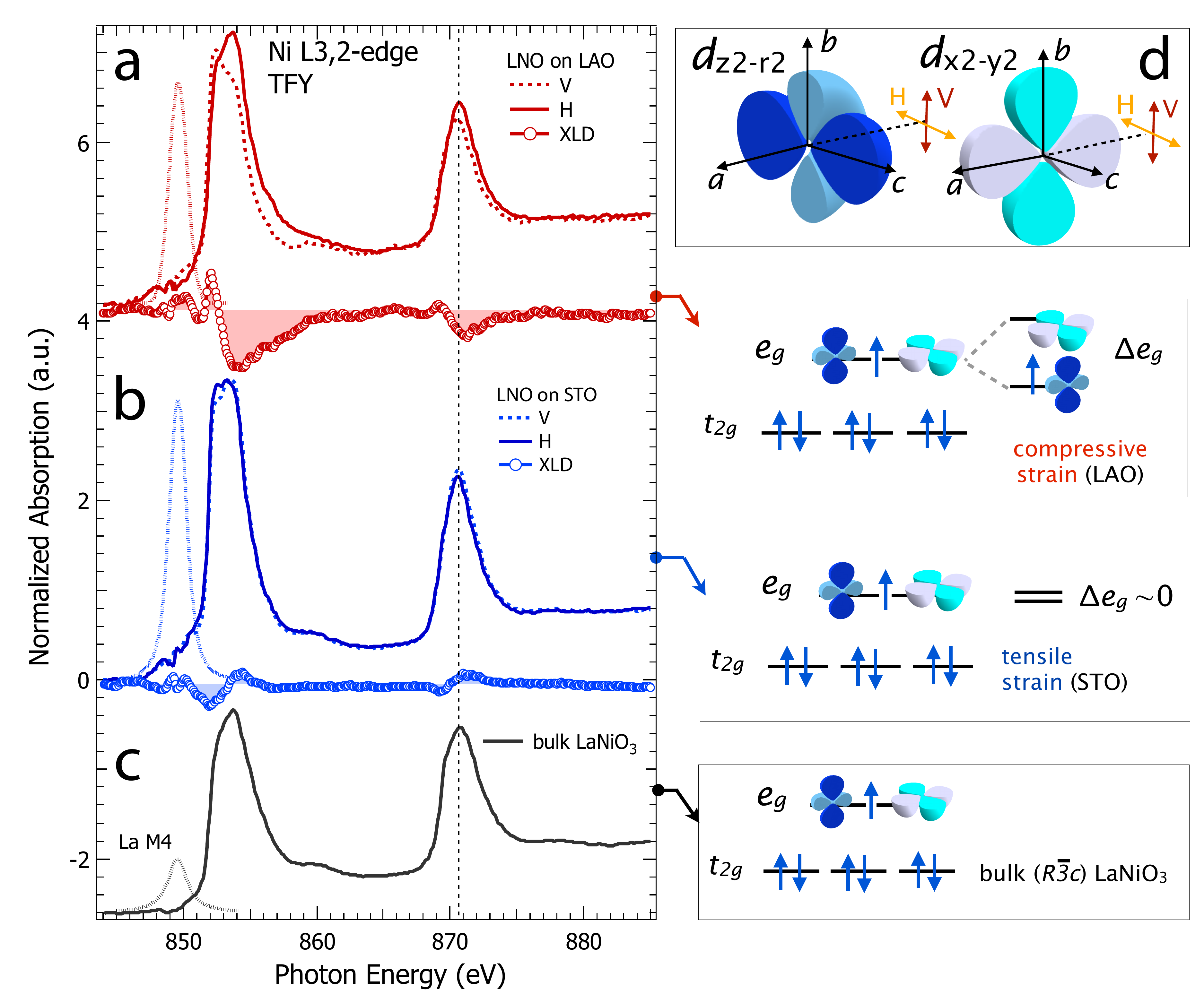}\vspace{-4pt}
   \caption{\label{Ni_XLD}(Color online)
   Normalized absorption spectra at the Ni L-edge
   (fluorescence mode) with varying
   photon polarization (see legend) for films
   under (a) compressive and (b) tensile 
   strain, on  LAO  and STO substrates, respectively, and 
   (c) bulk (ceramic) LNO.
   (\textsc{right}) Schematic orbital level diagram for bulk LNO and
   the inferred SIOP-effect on the $e_g$ doublet.}
\end{figure}

To explore the microscopic origins of the asymmetric strain-induced  
electronic states, we probe the local
occupied 3$d$ orbital configurations and the oxygen-derived 
states using both X--ray absorption spectroscopy (XAS) and X-ray linear dichroism 
(XLD) measurements at the Ni L- and O K-edges.
Figures~\ref{Ni_XLD}(a)-(c) show the normalized room temperature XAS spectra at the Ni L$_{3,2}$ edge for
polarization oriented parallel to $a - b$ vs.\ $c$ axes under different
epitaxial strain states.
Here, we focus on two representative films under (on) compressive (LAO) and 
tensile (STO) strain states (substrates).
Each polarization averaged spectra is representative of octahedrally
coordinated Ni$^{3+}$ in good agreement with the valence of bulk LNO.
There are distinct differences, however, in the polarization dependence
between films under compressive and tensile strain:
The absorption for photon polarization perpendicular to the
NiO$_2$ sheets is shifted $\sim$0.1~eV lower in energy than the
in-plane polarization ($a-b$) for LNO under compressive strain
on LAO [Fig.~\ref{Ni_XLD}(a)].
The dichroism indicates a small conduction band splitting $\Delta e_g$ between
the Ni-derived $d_{x^{2}-y^{2}}$ and $d_{3z^{2}-r^{2}}$ orbitals
(Fig.~\ref{Ni_XLD}, right) due to the distortion of the spherical charge distribution
around the Ni sites by non-uniform Ni--O bond lengths.
The sign of the XLD implies that the in-plane  Ni-O bonds are
moderately compressed relative to the out-of-plane apical Ni--O bonds
along the $c$-axis in agreement with both the model of \emph{symmetric} SIOP 
and the measured axial ratio for the films on LAO.
For STO with the same substrate--film orientation, but strain of the 
opposite sign, the SIOP model predicts an inversion of the orbital polarization 
and the population of the $d_{x^{2}-y^{2}}$ orbital; however,
consistent with our measured $R_H^{-1}$ and TEP, the polarization dependence 
of the films under tensile strain is remarkably different.
Fig.~\ref{Ni_XLD}(b) shows the orbital dichroism is absent, suggesting 
that uniform Ni--O bond lengths
and a highly symmetric cubic crystal field  ($\Delta e_g \sim 0$) remain 
seemingly intact despite the large ($\sim2\%$) tensile strain.
These findings are surprising as the 
in-plane lattice parameters are expanded by 2\% and 
due to coherent film registry with substrate, the symmetric SIOP should
have produced a  large orbital polarization.
To elucidate how the strain-stabilized film structure 
produces an asymmetric orbital polarization, 
we perform density functional calculations within
the local-spin density approximation (LSDA) plus Hubbard $U$ method 
following the details in Ref.\ \onlinecite{May/Rondinelli:2010}.
We carry out structural optimizations for homoepitaxially
strained nickelate films using our experimentally
determined lattice parameters.
We find that samples under compressive strain have an
atomic structure characterized by Ni--O--Ni bond angles, which are larger within the
epitaxial $xy$-plane than those along the growth ($z$-) direction.
The structure also consists of a single Ni$^{3+}$ site with different
in-plane and out-of-plane Ni--O bond lengths that lift the
formerly bulk $D_{3d}$ Ni-site symmetry.
Films under tensile strain  exhibit rotated octahedra 
but with an additional intriguing
\emph{breathing} distortion imposed on the
NiO$_6$ octahedra [Fig.~\ref{fig:RH_TEP}(inset)].
The breathing octahedral distortion
induces a Ni--O bond length disproportionation and splits 
the formerly bulk equivalent Ni sites into Ni(1) and Ni(2). They are 
located at the center of nearly ideal NiO$_6$ octahedra with average in-plane and
out-of-plane bonds lengths of 2.00~{\AA} and 1.93~\AA,
respectively.
Interestingly, an analogous tendency to bond disproportionation  leading to 
charge disproportionation  of the  type
Ni$^{(3\pm \delta)+}$ is present in the low-temperature monoclinic phases of 
small rare-earth nickelates \cite{Catalan:2008,Garcia-Munoz_et_al:2009}, 
but not for bulk LNO.
Note, the NiO$_6$ octahedral breathing distortion 
is unstable for compressive  strain ---a single Ni$^{3+}$ site is preferred.

\begin{figure}
\centering
   \includegraphics[width=0.98\columnwidth]{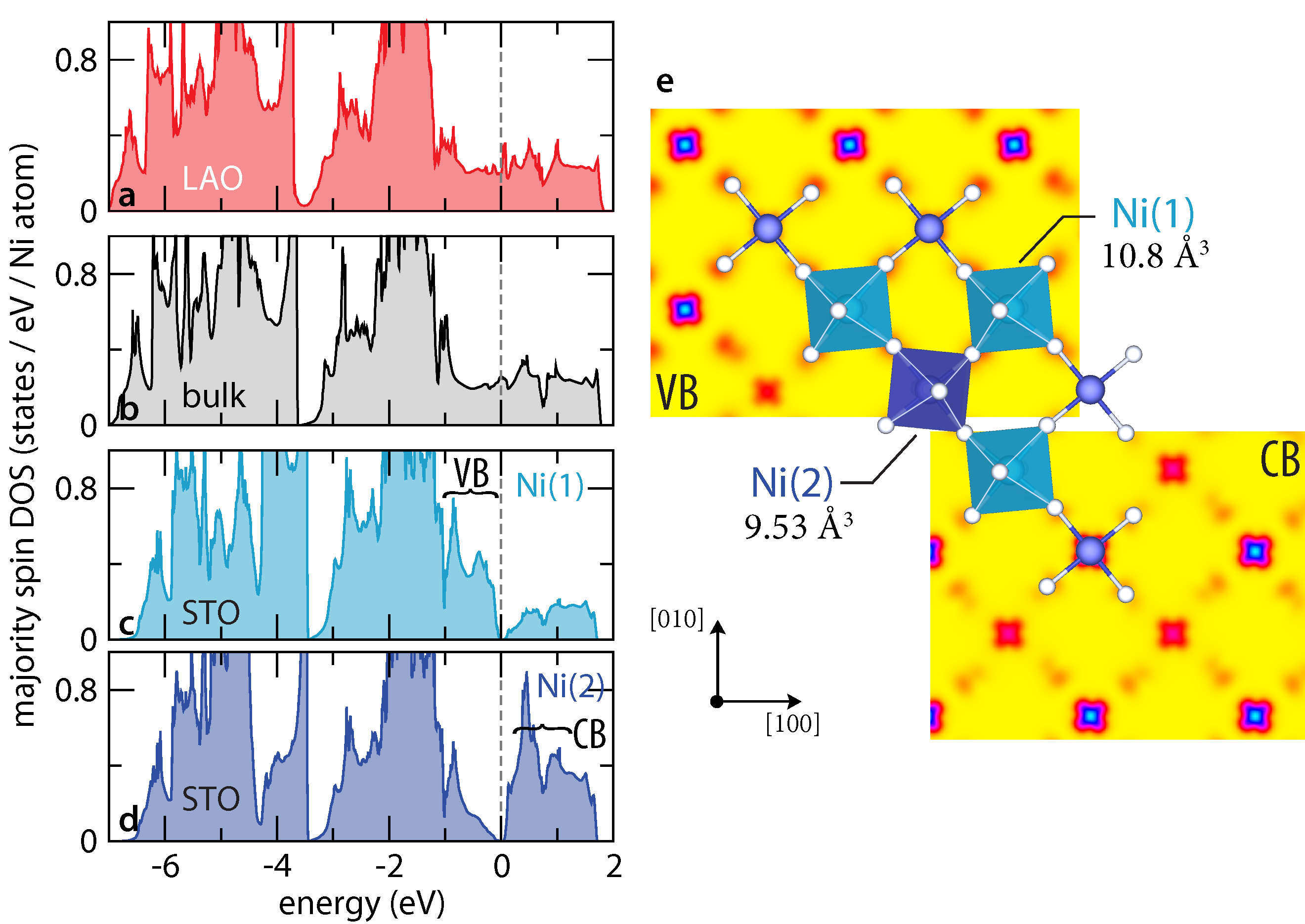}\vspace{-4pt}
   \caption{\label{fig:dos_composite}(Color online)
   The majority spin Ni 3$d$ density of states (DOS)
   for LNO  on (a), LAO substrate; for (b),
   a strain-free bulk crystal;
   and on (c) and (d), STO substrate.
   (e) The electron density for LNO on STO in the
   NiO$_2$ epitaxial plane [values integrated between $-$0.70 and
   0~eV (denoted as valence band, VB) and 0 and 1.7~eV (denoted as
   conduction band, CB)] reveals the 3D checkerboard spatial distribution of the Ni $e_g$
   states derived from the small and large octahedral units.
   [Darker (blue) areas indicate a higher value of the charge density.]}
\end{figure}
We next compare the  electronic ground state of  LNO on LAO and STO with the bulk  electronic
structure.
For films under compressive strain [Fig.~\ref{fig:dos_composite}(a)], we find a strain-induced
crystal field splitting of the Ni 3$d$ orbitals
with an electronic structure that remains  metallic, as in the bulk [Fig.~\ref{fig:dos_composite}(b)], 
albeit  with a strongly increased  bandwidth of $\sim$0.20~eV.
We find that the low-spin Ni$^{3+}$ electronic configuration is  accurately
described as itinerant and strongly mixed with an oxygen ligand hole, 
$d^{8}\underline{L}$.
In sharp contrast, films under tensile strain on STO show a semiconducting
gap of 0.10~eV with the bottom of the valence band edge
shifted by $\sim$0.5~eV over the compressively strained film [Fig.~\ref{fig:dos_composite}(c,d)].
We observe a modulation in the Ni--O bond covalency through the 
$dp\sigma$ band splitting, which produces an energy gap between the antibonding $e_g$ orbitals, 
that is stabilized by the emergent strain-induced octahedral breathing distortion not found in the bulk.
Figures~\ref{fig:dos_composite}(c) and (d) show  that the valence band of LNO under 
tensile strain is primarily composed of the $e_g$ states from the more ionic Ni(1)
cation with nearly one electron in each $d_{x^{2}-y^{2}}$ and $d_{3z^{2}-r^{2}}$ atomic orbital
($S=1$).
The suppressed intensity at the valence band edge indicates that this orbital occupation
is due to charge transfer  from the covalent  
Ni(2) site to the larger and more ionic Ni(1) site.
We calculate the charge transfer $\delta$ between Ni sites to be $\sim$0.20$e$
and homogeneously ordered throughout the film [Fig.~\ref{fig:dos_composite}(e)].
Because $\delta$ is small and the low-spin Ni$^{3+}$ valence is rare, it is equally likely 
that a bond-centered double ligand hole $\underline{\underline{L}}$  state, similar to 
the Zhang-Rice states in hole doped cuprates \cite{Zhang/Rice:1988}, could be stabilized over the 
site-centered configuration.
Nonetheless, our  XLD data and first-principles calculations show that for LNO  
films under tension, the preferred electronic configuration is one {\it without} a net 
$d$-orbital polarization.
The semi-ionic state results from a charge density modulation that occurs in the presence of 
orbital level splittings induced by tensile strain; in contrast, compressive 
strain favors a uniform orbital polarization of the Ni sites.

\begin{figure}
\centering
   \includegraphics[width=0.98\columnwidth]{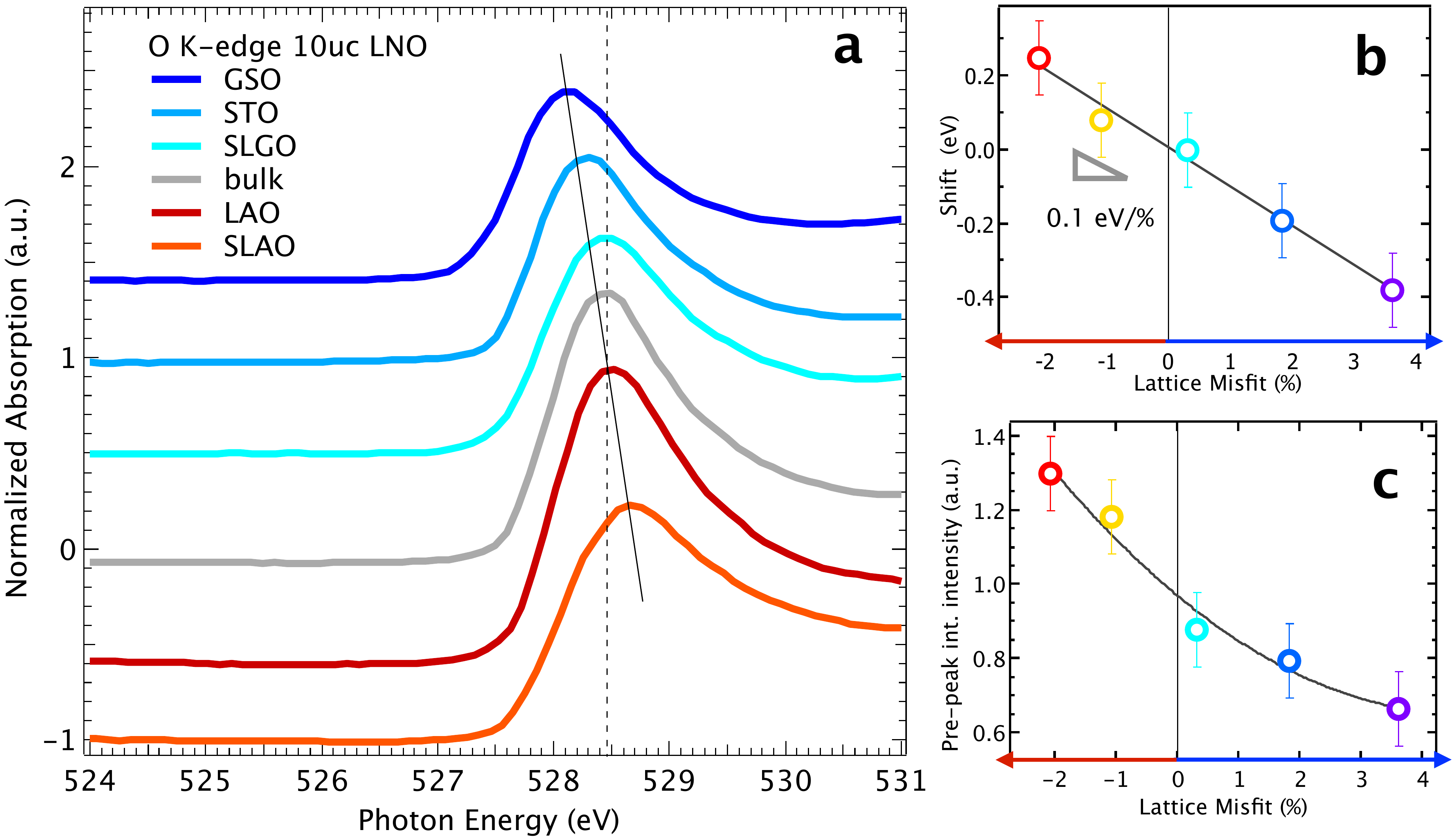}
   \caption{\label{fig:oxygen}(Color online) Bi-axial strain-induced covalency changes:
   (a) Low energy O K-edge pre-peak absorption around 528.5 eV for
   compressive (red) and a range of tensile (shades of blue)  strain states
   (a.u., arbitrary units).
   We omit peaks above 530 eV (transitions into La $5d4f$ and Ni $4s4p$ states) 
   for clarity.
   (b) Energy shift and
   (c) the normalized FWHM relative to the bulk LNO energy position with misfit strain.}
\end{figure}

To evaluate the  degree of strain-modulated covalency, we use resonant XAS measurements at the oxygen
K-edge to isolate the change in the ligand hole density of the $d^8\underline{L}$ component of the Ni$^{3+}$ ground state.
The oxygen spectrum around 528.5 eV  corresponds to transitions 
from the core-level O $1s$ state into the unoccupied and strongly hybridized
O $2p$ -- Ni 3$d$ states [Fig.~\ref{fig:oxygen}(a)]; the observed pre-peak intensity provides a
direct measure of the degree of  Ni--O bonding.
Unlike the Ni L-edge which is well aligned to the bulk LNO value,
the oxygen pre-peak reveals a remarkably systematic and nearly linear
dependence of the peak position with strain [Fig.~\ref{fig:oxygen}(b)]: 
it shifts at a rate of almost 100 meV
per percent lattice misfit, indicating direct strain control of the 
ligand hole density. 
Fig.~\ref{fig:oxygen}(c) shows a corresponding systematic change in
the full-width at half-maximum (FWHM), suggesting a strong narrowing of the 
electron bandwidth due to the deviation of the  Ni--O--Ni bond 
angle away from the ideal 180$^\circ$ with increasing tensile strain.
In charge transfer oxides like nickelates and cuprates the position of the O
K-edge pre-peak directly couples to the charge transfer energy, which is also 
highly susceptible to the symmetry of the cation coordination \cite{Alexander/Murphy_etal:1991}.
The strain-stabilized atomic distortions should therefore modify the Madelung energy. 
The constant nickel valence across all strain states explored allows us to 
probe the limits of the cohesive energy tunability with X-ray photoemission spectroscopy (XPS). 
Indeed, we find a controllable modulation of the O~1$s$ core level 
energy by $\sim0.29$~eV  between LNO on LAO and STO (0.16~eV per percent strain) due to the asymmetric 
orbital-lattice interactions in agreement with our XAS results and calculations \cite{Note1}.
These experimental findings  demonstrate  that strain control of the ligand
hole density is due to a major Madelung energy response.
We conjecture that by reversibly accessing these asymmetric orbital--lattice interactions control of 
the charge transfer energy and $d$-orbital polarization should enable 
access to latent metal--insulator phase transitions in other classes of 
correlated electron systems \cite{liu:2010}.

In summary, using a combination of resonant X--ray spectroscopies
and first-principles calculations, we report the discovery of a fundamental
asymmetry in the heteroepitaxial strain-induced orbital occupation response of  ultra-thin 
perovskite films with orbitally degenerate transition metal ions. We suggest that knowledge of this asymmetric orbital--lattice interaction is fundamental to the rational design of quantum materials with exotic correlated phases and enhanced critical temperatures.

The authors acknowledge discussions with D.\ Khomskii, G.\ Sawatzky, N.\ Spaldin, S.\ May,  A.\ Millis, G.\ Khalliulin, and O.\ Andersen. The NDSEG (JMR), DOD-ARO (grant no.\ 0402-17291) and NSF (grant no.\ DMR-0747808) 
(JC) supported this project. JLC acknowledges support from Research Corporation.
Work at the APS is supported by the U.S.\ DOE 
under grant no.\ DEAC02-06CH11357.


\end{document}